\documentclass[12pt,a4paper]{article}
\usepackage[latin1]{inputenc}
\usepackage[T1]{fontenc}
\usepackage{amsmath}
\usepackage{amsfonts}
\usepackage{amssymb}
\usepackage[english]{babel}
\usepackage[dvips]{graphicx,color}
\title{Casimir effect at finite temperature in a real scalar field theory}
\author{C.~Ccapa Ttira and C.~D.~Fosco
\\
{\normalsize\it Centro At\'omico Bariloche and Instituto Balseiro}\\
{\normalsize\it Comisi\'on Nacional de Energ\'\i a At\'omica}
{\normalsize\it R8402AGP Bariloche, Argentina.}
}
\begin{document}
\date{}
\maketitle
\begin{abstract}
\noindent 
We use a functional approach to evaluate the Casimir free energy for a
self-interacting scalar field in $d+1$ dimensions, satisfying Dirichlet
boundary conditions on two parallel planes. 

\noindent When the interaction is turned off, exact results for the free
energy in some particular cases may be found, as well as low and high
temperature expansions based on a duality relation that involves the
inverse temperature $\beta$ and the distance between the mirrors, $a$. 

\noindent For the interacting theory, we derive and implement two different
approaches.  The first one is a perturbative expansion built with a thermal
propagator that satisfies Dirichlet boundary conditions on the mirrors. The
second approach uses the exact finite-temperature generating functional as
a starting point.  In this sense, it allows one to include, for example,
non-perturbative thermal corrections into the Casimir calculation, in a
controlled way.

\noindent We present results for calculations performed using those two
approaches.  
\end{abstract}
\section{Introduction}\label{sec:intro}
Casimir and related effects, where quantum effects depend upon the
existence of boundary conditions for a quantum field, have been extensively
studied~\cite{rev}. Many different points of view and approaches to this
kind of problem have been followed, under quite different sets of
assumptions regarding the system; i.e., its intrinsic properties, and the
conditions under which one wants to evaluate the Casimir force.

\noindent In particular, there has been much interest in studying the
Casimir effect at a finite temperature ($T>0$), a study that can be
undertaken at different levels, distinguished by the way of taking into
account all the possible $T>0$ effects. One could, for example, include
thermal effects in the description of the matter on the mirrors (a very
active research topic~\cite{dispers}), or rather for the physical
description of the vacuum field~\cite{Lim},  or for both. 

In this article, we shall assume perfect (at any temperature) mirrors, with
thermal effects restricted to the vacuum field.

It has been noted that thermal and Casimir-like effects share some
similarities; this should hardly be surprising, since both may be regarded,
essentially, as finite-size effects, the former in the Euclidean time
interval $[0,\beta]$ (with periodicity/antiperiodicity conditions), and the
latter for a spacial coordinate with Dirichlet or Neumann conditions. 

Even though the nature of the boundary conditions is different, it should
be expected that, when both effects are present, interesting relations
(`dualities') will arise between the dependence on the inverse temperaure
$\beta = T^{-1}$ and the distance between the mirrors. 

In this article, we present an approach to the calculation of the Casimir
free energy for a scalar field at finite temperature in $d+1$ dimensions,
subject to Dirichlet boundary conditions, including self-interactions. This
approach is based on the use of a path-integral formulation, whereby the
$d+1$-dimensional problem is mapped to a dimensionally reduced one, with
fields living on the boundaries. 

For all the cases considered we analyze the high and low
temperature expansions, computing also perturbative corrections to the
Casimir free energy due to the non-interacting fields.

This work is organized as follows: in section~\ref{sec:method} we present
our approach, introducing conventions and definitions. In
section~\ref{sec:scalarfree} we deal with the calculations and the
corresponding results for a free real scalar field. Interactions are
introduced in~\ref{sec:scalarint} and, in~\ref{sec:concl}, we present our
conclusions.
\section{The method}\label{sec:method}
The main object we shall be interested in is the free energy $F(\beta,a)$
for a real scalar field $\varphi$ in $d+1$ spacetime dimensions, which is
subject to Dirichlet boundary conditions on two plane mirrors, separated by
a distance $a$ along the direction corresponding to the $x_d$ coordinate. 

In terms of the corresponding partition function ${\mathcal Z}(\beta,a)$,
$F(\beta,a)$ is given by:
\begin{equation}
F(\beta,a) \;=\; -\frac{1}{\beta} \, \ln {\mathcal Z}(\beta,a) \;, 
\end{equation}
while for ${\mathcal Z}(\beta,a)$ we shall use its standard functional
integral representation:
\begin{equation}
 {\mathcal Z}(\beta,a) \;=\; \int \big[{\mathcal D}\varphi\big] \,
e^{- S(\varphi)}\;,
\end{equation}
where $S$ is the Euclidean action at $T>0$; i.e., with an imaginary time
variable restricted to the $[0,\beta]$ interval~\cite{Kapusta:2006pm} and
periodic boundary conditions for the fields with respect to this coordinate
are implicitly assumed.

In this work, we consider an action $S$ such that \mbox{$S = S_0 + S_I$},
with the free part of the action, $S_0$, is given by:
\begin{equation}\label{eq:defs0}
	S_0(\varphi) \;=\; \frac{1}{2} \, \int_0^\beta  d\tau  \int d^dx
	\big(\partial_\mu \varphi \partial_\mu \varphi + m^2 \varphi^2
	\big) \;,
\end{equation}
while the interaction part, $S_I$, is given by:
\begin{equation}\label{eq:defsi}
 S_I(\varphi) \;=\;\frac{\lambda}{4!}  \int_0^\beta  d\tau  \int d^dx \,  
\big[ \varphi(\tau, {\mathbf x}) \big]^4 \;.
\end{equation}
We have used brackets in $\big[{\mathcal D}\varphi\big]$ to indicate that
the path-integral measure includes only those field configurations
satisfying Dirichlet boundary conditions on the locii of the mirrors, which
correspond to two parallel planes. Accordingly, we use a coordinate system
such that, if the Euclidean coordinates are denoted by
\mbox{$x=(x_0,x_1,\ldots,x_d)$} ($x_0 \equiv \tau$), the mirrors then
correspond to the regions: $x_d=0$ and $x_d=a$, and will be parametrized as
follows:
\begin{equation}
	x = (x_\parallel, 0) \;,\;\; x = (x_\parallel, a) \;, 
\end{equation}
where $x_\parallel=(\tau,\mathbf{x_{\parallel}})=(\tau,x_1,\dots,x_{d-1})$. 

Then we have:
\begin{equation}\label{eq:defmeasure}
 \big[{\mathcal D}\varphi\big] \;=\; {\mathcal D}\varphi \; 
\delta\big[\varphi(\tau,{\mathbf x}_\parallel,0)\big] \;
\delta\big[\varphi(\tau,{\mathbf x}_\parallel,a)\big] \;,
\end{equation}
where the $\delta$'s with braketed arguments are understood in the functional sense; for example:
\begin{equation}
 \delta\big[\varphi(\tau,{\mathbf x}_\parallel,a)\big] \;=\; 
\prod_{\tau,{\mathbf x}_\parallel} \,  \delta\big(\varphi(\tau,{\mathbf
x}_\parallel,a)\big) \;\;,
\end{equation}
while the ones on the right hand side are ordinary ones. Besides, we assume
that the ($\beta$-dependent) factor that comes from the integration over
the canonical momentum has been absorbed into the definition of ${\mathcal
D}\varphi$, so that performing the integral over $\varphi$ does indeed
reproduce the partition function (without any missing factors).

Following~\cite{Kardar:1997cu}, the functional $\delta$-functions are
exponentiated by means of two auxiliary fields, $\xi_1(x_\parallel)$ and 
$\xi_2(x_\parallel)$, living in $d$ spacetime dimensions:
\begin{equation}\label{eq:z00f}
	{\mathcal Z}(\beta,a) \;=\; \int {\mathcal D}\varphi \, {\mathcal
D}\xi_1 \, {\mathcal D}\xi_2 \; e^{-S(\varphi) \,+\, i\,\int d^{d+1}x
J_p(x) \varphi(x)}\,, \end{equation} where we introduced the singular
current:
\begin{equation}
J_p(x)\;\equiv\;  \delta (x_d) \xi_1(x_\parallel) \,+\, \delta (x_d- a)
\xi_2(x_\parallel) \,,  
\end{equation}
whose support is the region occupied by the mirrors.

Since we shall treat interactions in a perturbative approach, it is
convenient to deal first with the free theory, as a necessary starting
point to include the interactions afterwards.
\section{Free
theory}\label{sec:scalarfree} When the theory is free ($\lambda=0$), $S = S_0$,
and the integral over
$\varphi$ becomes Gaussian. The resulting partition function, denoted by
${\mathcal Z}^{(0)}(\beta,a)$ is then: \begin{equation}\label{eq:zint1}
{\mathcal Z}^{(0)}(\beta,a) \;=\; {\mathcal Z}^{(0)}(\beta) \,\times \,  
	\int {\mathcal D}\xi_1 {\mathcal D}\xi_2 \; e^{-S_p(\xi)} \;,
\end{equation}
where ${\mathcal Z}^{(0)}(\beta)$ is the (free) partition function in the
absence of the mirrors, and: 
\begin{equation}
	S_p(\xi) \;=\; \frac{1}{2} \,\int d^dx_\parallel  d^dy_\parallel  
	\xi_a(x_\parallel) \Omega_{ab}(x_\parallel,y_\parallel)
	\xi_b(y_\parallel) \;,
\end{equation}
where $a,b=1,2$, and 
\begin{equation}\label{eq:omega}
\Omega(x_{\parallel};y_{\parallel})=\left[ \begin{array}{cc}
\Delta(x_{\parallel},0;y_{\parallel},0) &
\Delta(x_{\parallel},0;y_{\parallel},a) \\
\Delta(x_{\parallel},a;y_{\parallel},0) &
\Delta(x_{\parallel},a;y_{\parallel},a) \\
\end{array} \right] \;,
\end{equation}
where $\Delta$ is the free, imaginary-time, propagator. It may be written
explicitly as follows:
\begin{equation}
\Delta(\tau_x,\mathbf{x};\tau_y,\mathbf{y})=\frac{1}{\beta}\sum_n \int
\frac{d^d\mathbf{k}}{(2\pi)^d} e^{i\omega_n(\tau_x\ - \tau_y)
+i\mathbf{k}(\mathbf{x}-\mathbf{y})}\,
\widetilde{\Delta}(\omega_n,\mathbf{k})\;,
\end{equation}
with \mbox{$\widetilde{\Delta}(\omega_n,{\mathbf k}_\parallel) =
(\omega_n^2 + {\mathcal E}^2({\mathbf k}_\parallel))^{-1}$}.  

\noindent We have used the notation \mbox{${\mathcal E}({\mathbf
k}_\parallel) = \sqrt{ {\mathbf k}_\parallel^2 + m^2}$}, and
\mbox{$\omega_n = \frac{2 \pi n}{\beta}$} denotes the Matsubara
frequencies.  $\Delta$ is the inverse, in the space of $\tau$-periodic
functions, of $K\equiv(-\partial^2 +m^2)$. 

Expression (\ref{eq:zint1}) allows one to extract from the free energy the
term that would correspond to a free field in the absence of mirrors,
$F^{(0)}$, plus another contribution, which we shall denote by $F^{(0)}_p$:
\begin{equation}
	F^{(0)}(\beta,a) \;=\; F^{(0)}(\beta) \,+\,F^{(0)}_p(\beta,a) \;,
\end{equation}
\begin{equation}
	F^{(0)}_p(\beta,a)\,=\, - \frac{1}{\beta} \, 
	\ln\Big[\frac{{\mathcal Z}^{(0)}(\beta,a)}{{\mathcal Z}^{(0)}(\beta)}\Big] \;,
\end{equation}
which clearly contains the Casimir effect information (including thermal
corrections). However, it also carries information that is usually
unwanted, associated to the ever-present self-energy of the mirrors.
Indeed, we see that it includes a divergent contribution to the free
energy, which can be neatly identified, for example, by noting that, 
when $a \to \infty$:
\begin{equation}
	\frac{{\mathcal Z}^{(0)}(\beta,a)}{{\mathcal Z}^{(0)}(\beta)}\,\to\,
	\big[{\mathcal Z}^{(0)}_m(\beta)\big]^2 \;, 
\end{equation}
where ${\mathcal Z}_m^{(0)}$ is the contribution corresponding to one
mirror:
\begin{eqnarray}
	{\mathcal Z}^{(0)}_m(\beta)&=& \int {\mathcal D}\xi_1 \; 
e^{-\frac{1}{2}\int d^dx_\parallel \xi_1(x_\parallel) \Delta(x_{\parallel},0;y_{\parallel},0)
\xi_1(y_\parallel)} \nonumber\\
&=& \int {\mathcal D}\xi_2 \; e^{-\frac{1}{2}\int d^dx_\parallel \xi_2(x_\parallel) 
\Delta(x_{\parallel},a;y_{\parallel},a) \xi_2(y_\parallel)} \;.
\end{eqnarray}
We then identify
\begin{equation}
	F^{(0)}_m(\beta) \;\equiv\; -\frac{1}{\beta} 
	\ln\big[{\mathcal Z}^{(0)}_m(\beta)\big]
\end{equation}
as the free energy term that measures a mirror's self-interaction.  It is,
of course, independent of $a$. Extracting also this contribution, we have
the following decomposition for $F^{(0)}$ 
\begin{equation}\label{eq:decomp}
F^{(0)}(\beta,a)\;=\; F^{(0)}(\beta) \,+\, 2\,  F^{(0)}_m(\beta) \,+\, 
F^{(0)}_c(\beta,a)
\end{equation}
where $F^{(0)}_c(\beta,a)$ has a vanishing limit when \mbox{$a \to \infty$}.

We can produce a more explicit expression for the interesting term
$F^{(0)}_c$, starting from its defining properties above. Indeed, we have:
\begin{equation}
	F^{(0)}_c(\beta,a) \,=\, -\frac{1}{\beta} \,
	\ln \Big\{\frac{{\mathcal Z}^{(0)}(\beta,a)}{{\mathcal Z}^{(0)}(\beta)
	\big[{\mathcal Z}^{(0)}_m(\beta)\big]^2 }\Big\}\;, 
\end{equation}
which, by integrating out the auxiliary fields, may be written in terms of
the matrix kernel $\Omega$:
\begin{equation}
	F^{(0)}_c(\beta,a) \,=\, \frac{1}{2 \beta} {\rm Tr} \ln \Omega 
	\,-\, \frac{1}{2 \beta} {\rm Tr} \ln \Omega_\infty 
\end{equation}
where $\Omega_\infty \equiv\Omega|_{a \to \infty}$ and the trace is over
both spacetime coordinates and $a, b$ indices.

Assuming that an UV regularization is introduced in order to make sense of
each one of the traces above, and using `reg' to denote UV
regularized objects, we see that:
\begin{eqnarray}
	F^{(0)}_{c,reg}(\beta,a) &=& \frac{1}{2 \beta} \Big[{\rm Tr} \ln
	\Omega \Big]_{reg}
	\,-\, \frac{1}{2 \beta} \Big[{\rm Tr} \ln \Omega_\infty
	\Big]_{reg} \nonumber\\
	&=& \frac{1}{2 \beta} \Big[{\rm Tr} \ln\big(\Omega_\infty^{-1}
	\Omega \big) \Big]_{reg} \;.
\end{eqnarray}
As we shall see, the trace on the second line is finite, and has a
finite limit when the regulator is removed. Using some algebra we may reduce
the trace to one where only continuous indices appear:
\begin{equation}
	F^{(0)}_c(\beta,a) \,=\, \frac{1}{2\beta} \,{\rm Tr} 
	\ln \Omega_c \;, 
\end{equation}
where a `reduced' kernel $\Omega_c$ which is given by:
\begin{equation}
\Omega_c (x_\parallel,y_\parallel) \,=\, \delta(x_\parallel - y_\parallel)
\,-\, T(x_\parallel,y_\parallel) \;,
\end{equation}
with:
\begin{eqnarray}
	T(x_\parallel,y_\parallel)&=&	\int d^dz_\parallel  d^dw_\parallel d^du_\parallel
	[\Omega_{11}]^{-1}(x_\parallel,z_\parallel)
	\Omega_{12}(z_\parallel,w_\parallel)\nonumber\\
	&\times&
	[\Omega_{22}]^{-1}(w_\parallel,u_\parallel)\Omega_{21}(u_\parallel,y_\parallel)
	\;,
\end{eqnarray}
has been introduced.

Finally note that, due to translation invariance along the
mirrors, the free energy shall be
proportial to $V_p$, the area of the mirrors. Then, to absorbe this
divergence we shall rather consider the corresponding free energy {\em
density\/}, obtained by dividing the extensive quantity by $V_p$. In
particular,
\begin{equation}
	{\mathcal F}^{(0)}_c(\beta,a) \,\equiv \, \lim_{V_p \to \infty}
	\Big[\frac{1}{V_p} F^{(0)}_{c,V_p}(\beta,a)\Big] 
\end{equation}
where $ F^{(0)}_{c,V_p}$ on the right hand side is evaluated for a system with a
finite parallel volume. Moreover, taking advantage of the translation
invariance along the parallel coordinates, we may use a Fourier
transformation to write:
\begin{equation}\label{eq:denercas}
	{\mathcal F}^{(0)}_c(\beta,a) \;=\; \frac{1}{2 \beta}
	\int \frac{d^{d-1}{\mathbf k}_\parallel}{(2\pi)^{d-1}}  
	\sum_{n =-\infty}^{+\infty} 
	\ln \widetilde{\Omega_c}^{(n)}({\mathbf k}_\parallel)
	\;,
\end{equation}
where
\begin{equation}
	\widetilde{\Omega_c}^{(n)}(|{\mathbf k}_\parallel|)
	\;=\; 1 \,-\, \widetilde{T}^{(n)}({\mathbf k}_\parallel)\;,
\end{equation}
with:
\begin{equation}\label{eq:defT}
	\widetilde{T}^{(n)}({\mathbf k}_\parallel) \;\equiv\; 
	[{\widetilde\Omega}_{11}^{(n)}({\mathbf k}_\parallel) ]^{-1}
	\,{\widetilde\Omega}_{12}^{(n)}({\mathbf k}_\parallel ) \,
	[ {\widetilde\Omega}_{22}^{(n)}({\mathbf k}_\parallel )]^{-1} \,
	{\widetilde\Omega}_{21}^{(n)}({\mathbf k}_\parallel )\;.
\end{equation}
where the tilde denotes Fourier transformation in both $\tau$ and ${\mathbf
x_\parallel}$. Note the appearance of the reciprocals of the matrix
elements of ${\widetilde \Omega}^{(n)}$ (not to be confused with the matrix
elements of the inverse of that matrix).  The explicit form of the objects
entering in (\ref{eq:defT}) may be obtained from (\ref{eq:omega}):  
\begin{eqnarray}
{\widetilde\Omega}_{11}^{(n)}({\mathbf k}_\parallel) 
= {\widetilde\Omega}_{22}^{(n)}({\mathbf k}_\parallel) &=& 
\frac{1}{2 \sqrt{\omega_n^2 + {\mathcal E}^2({\mathbf k}_\parallel)}} \nonumber\\
{\widetilde\Omega}_{12}^{(n)}({\mathbf k}_\parallel) = 
{\widetilde\Omega}_{21}^{(n)}({\mathbf k}_\parallel) &=& 
\frac{e^{-  a\, \sqrt{\omega_n^2 + 
{\mathcal E}^2({\mathbf k}_\parallel)}}}{2 \sqrt{\omega_n^2 + {\mathcal
E}({\mathbf k}_\parallel)}} \;,
\end{eqnarray}
so that:
\begin{equation}
\widetilde{\Omega}_c^{(n)}({\mathbf k}_\parallel) \;=\; 1 \,-\, e^{- 2 a  \sqrt{\omega_n^2 + 
{\mathcal E}^2({\mathbf k}_\parallel)}} \;.
\end{equation}

It is then evident that the sum and the integral in (\ref{eq:denercas})
converge, since the integrand falls off exponentially for large values of
the momenta/indices. This behaviour should be expected, since we have
constructed this object subtracting explicitly the would-be self-energy
parts, the possible source of UV divergences.

Thus, the main result of the previous calculations is a (finite) expression
for the free energy, which can be written as follows:
\begin{equation}\label{eq:freexact}
	{\mathcal F}^{(0)}_c(\beta,a) \;=\; \frac{1}{2 \beta}
	\int \frac{d^{d-1}{\mathbf k}_\parallel}{(2\pi)^{d-1}}  
	\sum_{n =-\infty}^{+\infty} 
	\ln \Big[ 1 \,-\, e^{- 2 a  \sqrt{\omega_n^2 + {\mathcal
	E}^2({\mathbf k}_\parallel)}} \Big] \;.
\end{equation}
It is worth noting that the expression above has been obtained subtracting
the free energy corresponding to a situation where the mirrors are
separated by an infinite distance; on the other hand, we note that, when
the $\beta \to \infty$ limit is taken, the sum is replaced by an integral,
and we obtain:
\begin{equation}\label{eq:enexact}
	{\mathcal E}^{(0)}_c(a) \,\equiv\,{\mathcal F}^{(0)}_c(\infty,a) \,=\, 
	\frac{1}{2} \int \frac{d^dk_\parallel}{(2\pi)^d}  
	\ln \Big( 1 \,-\, e^{- 2 a  \sqrt{ k_\parallel^2 + m^2 }} \Big) \;,
\end{equation}
which is the Casimir energy per unit area.

With this in mind, one may introduce yet another quantity; the {\em
temperature dependent} part of the free energy, whose area density we will
denote by ${\mathcal F}_t^{(0)}(\beta,a)$, and, from the discussion above,
it is given by:
\begin{equation}
{\mathcal F}_t^{(0)}(\beta,a) = {\mathcal F}_c^{(0)}(\beta,a) \,-\, 
{\mathcal F}_c^{(0)}(\infty,a)\;.
\end{equation}
By using a rather simple rescaling it the integral in the second term, we may write:
\begin{eqnarray}
{\mathcal F}_t^{(0)}(\beta,a) &=& \frac{1}{2 \beta} 
\int \frac{d^{d-1}{\mathbf k}_\parallel}{(2\pi)^{d-1}} 
\Big\{ 
\sum_{n =-\infty}^{+\infty} \ln\big[ 1 \,-\, e^{ - \frac{2 a}{\beta}
\sqrt{(2 \pi n)^2 + (\beta {\mathcal E}({\mathbf k}_\parallel))^2 }}\big]
\nonumber\\
&-& \int_{-\infty}^\infty  d\nu \, \ln\big[ 1 \,-\, 
e^{ - \frac{2 a}{\beta} \sqrt{(2 \pi \nu)^2 + (\beta {\mathcal E}({\mathbf
k}_\parallel))^2}}\big] \Big\} \;,
\end{eqnarray}
which yields the Casimir free energy as a difference between a series and
an integral, although the for the thermal part, since the sum over energy
modes has already been (implicitly) performed.
 
Before evaluating the free energy for different numbers of spacetime
dimensions, we explore below some consequences of a duality between $\beta$
and $a$.
\subsection{Duality}\label{ssec:dual}
The dual role played by $\beta$ and $a$ in the free energy may be
understood, for example, by attempting to calculate that object by
following an alternative approach, based on the knowledge of the exact
energies of the field modes, emerging from the existence of Dirichlet
boundary conditions. 

The energies $\omega_l$ of the stationary modes are:
\begin{equation}
w_l(\mathbf{k_\parallel})\,=\,\sqrt{\frac{\pi^2
l^2}{a^2}+\mathbf{k}_\parallel^2 + m^2}\;\;, \;\;\;\; l \in {\mathbb N}\;.
\end{equation}

Since each one of these modes behaves as a harmonic oscillator degree of
freedom, its free energy $f\big[w_l({\mathbf k}_\parallel)\big]$ has the
following form:
\begin{eqnarray}
	f\big[w_l({\mathbf k}_\parallel)\big]&=& 
	-\frac{1}{\beta} \,\ln \big[ \sum_{N=0}^\infty 
	e^{-\beta w_l({\mathbf k}_\parallel) (N+\frac{1}{2})} \big] \nonumber\\
	&=& \frac{1}{2} w_l({\mathbf k}_\parallel)  + \frac{1}{\beta} \, 
	\ln \big[ 1 \, - \,  e^{-\beta w_l({\mathbf k}_\parallel)} \big]\;.
\end{eqnarray}
Now we consider ${\mathcal F}_t^{(0)}(\beta,a)$, the {\em temperature
dependent\/} part of the free energy density, obtained by summing the
second term in the expression above over all the degrees of freedom, and
dividing by the parallel area: \begin{equation}
	{\mathcal F}_t^{(0)}(\beta,a)\,=\, \frac{1}{\beta} \, 
	\int \frac{d^{d-1}{\mathbf k}_\parallel}{(2\pi)^{d-1}}  
	\sum_{l = 1}^{+\infty} \ln \big[ 1 \, - \,  
	e^{-\beta w_l({\mathbf k}_\parallel)} \big] \;.
\end{equation}
This may also be written as follows:
\begin{eqnarray}\label{eq:freet}
	{\mathcal F}_t^{(0)}(\beta,a) &=& \frac{1}{2 \beta} \, \int
	\frac{d^{d-1}{\mathbf k}_\parallel}{(2\pi)^{d-1}}  
	\sum_{l = - \infty}^{+\infty} \ln \big[ 1 \, - \,  e^{-\beta w_l({\mathbf
	k}_\parallel)} \big] \nonumber\\ 
	 &-& \frac{1}{\beta} \, \int \frac{d^{d-1}{\mathbf k}_\parallel}{(2\pi)^{d-1}}  
	 \ln \big[ 1 \, - \,  e^{-\beta {\mathcal E}({\mathbf k}_\parallel)} \big] \;.
\end{eqnarray}
We now recall that we are considering the free energy for one and the same system,
albeit with different normalizations: subtraction at $a \to \infty$ in
(\ref{eq:freexact}), and at $\beta \to \infty$ in (\ref{eq:freet}). Then we
may write:
\begin{equation}
	{\mathcal F}_c^{(0)}(\beta,a) \,-\, \lim_{\beta \to \infty} 
	{\mathcal F}_c^{(0)}(\beta, a) \,=\, 
	{\mathcal F}_t^{(0)}(\beta,a) \,-\, \lim_{a \to \infty} 
	{\mathcal F}_t^{(0)}(\beta, a) \;, 
\end{equation}
or:
\begin{equation}
	{\mathcal F}_c^{(0)}(\beta,a) \,-\, {\mathcal E}_c^{(0)}(a) \,=\, 
	{\mathcal F}_t^{(0)}(\beta,a) \,-\, {\mathcal F}_t^{(0)}(\beta) \;, 
\end{equation}
where ${\mathcal F}_t^{(0)}(\beta)$ is an $a$-independent object, which
corresponds to the free energy of a free field.

Thus, if we are going to be concerned with temperature and $a$-dependent
quantities, for example, to study the temperature dependence of the Casimir
force, we only need derivatives with respect to $\beta$ and $a$ of the
expression above, and we see that:
\begin{equation}
	\frac{\partial^2}{\partial a  \partial \beta} \big[{\mathcal
	F}_c^{(0)}(\beta,a) \big]
	\,=\,\frac{\partial^2}{\partial \beta  \partial a} 
	\big[ {\mathcal F}_t^{(0)}(\beta,a) \big]\;, 
\end{equation}
or:
\begin{equation}\label{eq:aux1}
	\widetilde{\mathcal F}_c^{(0)}(\beta,a) \,=\, 
	\widetilde{\mathcal F}_t^{(0)}(\beta,a) \, \equiv \,	
	\widetilde{\mathcal F}^{(0)}(\beta,a)  \;, 
\end{equation}
where the tildes denote subtraction of any term which has a vanishing mixed
second partial derivative.

On the other hand, from the explicit form of the free energy density as sum
over modes (Matsubara or Casimir), we find the identity:
\begin{equation}
\big[\beta \widetilde{\mathcal F}_t^{(0)}\big] (2 a , \beta/2) \,=\, 
\big[ \beta \widetilde{\mathcal F}_c^{(0)}\big] (\beta,a) \;, 
\end{equation}
which, combined with (\ref{eq:aux1}) yields a duality between $a$ and $\beta$:
\begin{equation}\label{eq:duality}
	\widetilde{\mathcal F}^{(0)}(2 a , \beta/2) \,=\, \frac{\beta}{2 a}
	\, \widetilde{\mathcal F}^{(0)}(\beta,a)\;. 
\end{equation}
To proceed, we make the simplifying assumption that $m=0$, and transform
variables in the integral over $\mathbf{k_\parallel}$ to have a
dimensionless integral, obtaining
\begin{equation}
\beta \,\widetilde{\mathcal F}^{(0)}(\beta,a)\;=\;(\frac{2\pi}{\beta})^{d-1} g(\gamma,d) \,,
\end{equation}
where $g(\gamma,d)$ is given by
\begin{equation}\label{eq:G}
g(\gamma,d)\,=\,C_d \,\sum_n \int_0^{\infty} dx\, x^{d-2}\,\ln
\left(1-e^{-\gamma \sqrt{n^2+x^2}} \right) \;,
\end{equation}
where $\gamma= 4\pi a/\beta$ and $C_d$ is a constant factor that depends solely on the dimension $d$.

Now, using the duality formula ($\ref{eq:duality}$), we obtain the an
interesting relation involving $g$:
\begin{equation}\label{eq:Tdualidad}
\gamma^{d-1}g(\gamma,d)\,=\,\alpha^{d-1}g(\alpha,d)\,,
\end{equation}
where $\alpha=\beta \pi /a$  (or $\alpha = (2 \pi)^2/\gamma$). Note that
this relation has immediate relevance to relate the low and high
temperature regimes. Indeed, writing the expression above more explicitly,
we see that:
\begin{equation}\label{eq:Tdualidadp}
	g(\frac{4 \pi}{\beta} , d)\,=\, \big(\frac{\beta}{2 a}\big)^{d-1} \,
	g(\frac{\pi \beta}{a} , d)\,.
\end{equation}

This duality relation had been pointed out, for the $d=3$ case,  by Balian
and Duplantier~\cite{balian:2004}. In the remainder of this section, we
apply the previous results and study some particular properties of the free
energy corresponding the the free scalar field, firstly for the $d=1$ case,
which we single out since it can be exactly solved, and then for $d > 1$.  

\subsection{$d=1$}
The free energy (neglecting an irrelevant constant) is given by the expression:
\begin{equation}\label{eq:ecas1+1}
 {\mathcal F_c^{(0)}}(\beta,a)= \frac{1}{2 \beta} \sum_{n\ne0} \, \ln
 (1-e^{- 2 a \mid \omega_n \mid})\,,
\end{equation}
which may be written as an infinite product:
\begin{equation}\label{eq:ecas1+1Tb}
  {\mathcal F_c^{(0)}}(\beta,a)= \frac{1}{\beta} \, \ln \prod_{n=1}^{+\infty}(1-q^{2n}),
\end{equation}
where $q=e^{-2 \pi a /\beta}$,  and $\mid q \mid$ <1 for $T>0$. Using 
standard properties of ellipthic functions, we get a more 
explicit results for the free energy:
\begin{equation}\label{eq:defzz0}
  {\mathcal F_c^{(0)}}(\beta,a)= \frac{\pi a}{6 \beta^2}+ \frac{1}{\beta} \, \ln[\eta(2a/\beta \,i)],
\end{equation}
where $\eta(z)$  is Dedekind's eta function. Although it has been obtained
for $T>0$, one can verify that it also yields the proper results for 
\mbox{$T \to 0$}, namely, 
\mbox{${\mathcal E}_c^{(0)}(a)= -\frac{\pi}{24 a}$}.

Besides, the $\eta$-function satisfies the property: 
\mbox{$\eta (1/z) = \sqrt{i z}\, \eta(z)$} which, in this context, 
is tantamount to the duality relation. 

\subsection{$d > 1$}
Now, we exploit the duality relation to extract the high and low
temperature behavior of the Casimir energy, for $d>1$. In a high
temperature expansion, formula ($\ref{eq:G}$) can be expanded for small
$e^{-\gamma}$. The infinite-temperature limit corresponds to $\gamma \to
\infty$, thus $n=0$ yields the leading contribution in this expansion, with
the $n=1,\,2, \ldots $ terms producing higher-order corrections.

The explicit form of the leading high temperature term is:
\begin{equation}\label{eq:Ftt}
	\tilde{\mathcal F}_c^{(0)}(\beta,a) \,\sim\,- 
\frac{1}{2\pi^{d/2} \beta}\frac{\Gamma(d/2)\zeta(d)}{(2a)^{d-1}} \;,\;\;
\beta \sim 0\;,
\end{equation}
whereas aplying the duality formula (\ref{eq:duality}) $a \to \beta /2$, 
we have the first order contribution in the low temperature regime,
\begin{equation}\label{eq:Ft0}
{\mathcal F_c^{(0)}}(2a,\beta/2)\,\sim\,-\frac{1}{2\pi^{d/2}}
\frac{\Gamma(d/2)\zeta(d)}{(\beta)^d}\;\;\; (\beta \to \infty),
\end{equation}
which is $a$ independent. The sub-leading contributions ($n=1,2,...$)
include more involved integrals; nevertheless, when $d=3$ and $\gamma \to
\infty$, it reduces to terms like $e^{-n \,\gamma }$. Thus, the most
significant contribution for $T\to \infty$ is:  
\begin{equation}\label{eq:Ftt1}
-\frac{1}{2 a \beta^2} e^{-4\pi a /\beta}\,.
\end{equation}
Therefore, the duality formula implies that, when $T \to 0$,
\begin{equation}\label{eq:Ft01}
\tilde{\mathcal F}_c^{(0)}(\beta,a) \sim -\frac{1}{2 a \beta^2} e^{-\pi \beta /a}\,.
\end{equation}
Collecting these results for $d=3$, we see that, for high temperatures:
$T \to \infty$
\begin{equation}
{\mathcal F_c^{(0)}}(\beta,a)\,\sim\,-\frac{1}{16 \pi} 
\frac{\zeta(3)}{a^2 \beta}-\frac{1}{2 a \beta^2} e^{-4\pi a /\beta}\,\,,
\end{equation}
while, for low temperatures the corresponding behaviour is: 
\begin{equation}
{\mathcal F_c^{(0)}}(2a, \beta/2)\,
\sim\,-\frac{\pi^2}{1440 a^3} \left[1+ \frac{360}{\pi^3}\frac{a^3}{\beta^3}\zeta(3)+ 
\frac{720}{\pi^2}\frac{a^2}{\beta^2} e^{-\pi \beta /a} \right]\,.
\end{equation}
\section{Interacting theory}\label{sec:scalarint} 
Let us now consider the theory that results from turning on the quartic
self-interaction term, studying the corresponding corrections to the free
energy. 

This can be done in (at least) two different ways. We shall consider first
the would-be more straightforward approach, which amounts to expanding the
interaction term in powers of the coupling constant first, postponing the
integration over the scalar field until the end of the calculation. We call it
`perturbative' since it is closer in spirit to standard perturbation theory.

\subsection{Perturbative approach}\label{ssec:pert}
We have to evaluate:
\begin{eqnarray}
{\mathcal Z}(\beta, a) &=& \int [{\mathcal D}\varphi] \,e^{-S_I(\varphi)}
\,  e^{-S_0(\varphi)} \nonumber\\
 &=& {\mathcal Z}^{(0)}(\beta, a) \;\; \langle e^{-S_I(\varphi)} \rangle \;, 
\end{eqnarray}
where we have used $\langle \ldots \rangle$ to denote functional averaging
with the free (Matsubara) action and the scalar-field integration measure satisfying
Dirichlet boundary conditions. Namely,
\begin{equation}
\langle \ldots \rangle \;\equiv\; 
\frac{\int [{\mathcal D}\varphi] \ldots e^{-S_0(\varphi)}}{\int [{\mathcal
D}\varphi] \, e^{-S_0(\varphi)}} \;.
\end{equation}

Thus, the different terms in the perturbative expansion are obtained by
expanding in powers of the coupling constant, $\lambda$.   For example, the
first order term, ${\mathcal F}^{(1)}$, may be written as follows:
\begin{equation}
{\mathcal F}^{(1)}(\beta,a) \;=\; \frac{\lambda}{4! \beta} \int 
d^{d+1}x \, \langle [\varphi(x)]^4 \rangle \;,
\end{equation}
which may be expressed (via Wick's theorem), in terms of the average
of just two fields. Indeed, the generating functional for these
averages,  $Z_0^J(\beta,a)$,  is simply:
\begin{equation}\label{eq:zJcomplete}
 {\mathcal Z}_0^J(\beta,a)\;=\; \int \big[{\mathcal D}\varphi\big] \, e^{-
 S_0(\varphi) + \int d^{d+1}x \,  J(x) \varphi(x)}\;.
\end{equation}
We again introduce the auxiliary fields to impose the periodicity
constraints on the measure, so that: 
\begin{equation}\label{eq:genfunct0}
	{\mathcal Z}_0^J(\beta,a) \,=\, 
\int {\mathcal D}\varphi {\mathcal D}\xi \,
	\, e^{ - S_0[\varphi] + \int d^{d+1}x [i J_p(x)+J(x)]\varphi (x)}\,.
\end{equation}
Integrating out the scalar field $\varphi$, and using a simplified notation
for the integrals, we obtain:
\begin{eqnarray}
	{\mathcal Z}_0^J(\beta,a) &=& {\mathcal Z}_0(\beta,a) \, \int {\mathcal D}\xi \,
	\exp \Big[ - \frac{1}{2} \int_{x_\parallel,y_\parallel} 
	\xi_\alpha(x_\parallel)
	\Omega_{\alpha\beta}(x_\parallel,y_\parallel) \xi_\beta(y_\parallel) \nonumber\\
	&+& i \int_{x_\parallel}\xi_\alpha(x_\parallel) L_\alpha(x_\parallel) \,+\, 
	\frac{1}{2} \int_{x,y} J(x) \Delta(x,y) J(y) \Big] \;,
\end{eqnarray}
where: $L_\alpha (x_\parallel) \equiv \int_y \Delta(x_\parallel,a_\alpha;y)
J(y)$, with $a_\alpha = (\alpha - 1) a$, $\alpha=1,2$.
Integrating now the auxiliary fields, we see that:
\begin{equation}
	{\mathcal Z}_0^J(\beta,a) \;=\; {\mathcal Z}_0 (\beta,a)\, 
	e^{\frac{1}{2} \int_{x,y} J(x) G(x,y) J(y) } \;,
\end{equation}
where $G$, the (free) thermal correlation function in the presence of the
mirrors, may be written more explicitly as follows:
\begin{eqnarray}
G(x;y) \;= \; \langle \varphi(x) \varphi(y) \rangle &=& \Delta(x;y) \;-\;
\int_{x'_\parallel,y'_\parallel} \Big\{ \Delta(x_\parallel,x_d;
x'_\parallel,a_\alpha) \nonumber\\  
&\times& [\Omega^{-1}]_{\alpha\beta}(x'_\parallel,y'_\parallel) 
\Delta(y'_\parallel,a_\beta; y_\parallel,y_d) \Big\} \;,
\end{eqnarray}
and this is the building block for all the perturbative corrections.

It is quite straightforward to check that the propagator above does verify
the Dirichlet boundary conditions, when each one of its arguments approaches
a mirror. For example:
\begin{eqnarray}
\lim_{x_d \to a_\gamma} G(x;y) &=& \Delta(x_\parallel,a_\gamma;y) \;-\;
\int_{x'_\parallel,y'_\parallel} \Big\{ \Omega_{\gamma \alpha}(x_\parallel; x'_\parallel) 
[\Omega^{-1}]_{\alpha\beta}(x'_\parallel,y'_\parallel) 
\Delta(y'_\parallel,a_\beta; y_\parallel,y_d) \Big\} \nonumber\\
 &=& \Delta(x_\parallel,a_\gamma;y)  \;-\; \int_{y'_\parallel}
 \delta_{\gamma \beta} \delta(x_\parallel,y'_\parallel) 
\Delta(y'_\parallel,a_\beta; y_\parallel,y_d) \nonumber\\
&=& \Delta(x_\parallel,a_\gamma;y) \,-\, \Delta(x_\parallel,a_\gamma;y)
\;=\;0 \;,
\end{eqnarray}
and analogously for the second argument.

The thermal correlation function $G(x;y)$ above is composed of two contributions: 
one is the usual free thermal propagator in the absence of boundary conditions and
the other come from reflexions on the surface boundaries. In this way we
write 
\begin{equation}
G(x;y)\,=\, \Delta(x;y)-M(x;y)\,,
\end{equation}
where $M(x;y)$ is given by:
\begin{equation}\label{eq:Mpropagador}
M(x;y)=\frac{1}{\beta}\sum_n \int \frac{d^{d-1}\mathbf{k}_\parallel}{(2\pi)^{d-1}} e^{iw_n(\tau_x\ - \tau_y)+i\mathbf{k}_\parallel(\mathbf{x}_\parallel-\mathbf{y}_\parallel)}\,
\widetilde{M}(w_n,\mathbf{k}_\parallel;x_d,y_d)\,\,,
\end{equation}
with $\widetilde{M}(w_n,\mathbf{k}_\parallel;x_d,y_d)$ equal to:
\begin{equation}\label{eq:Mpropagfourier}
\frac{e^{-(|x_d|+|y_d|)E}-e^{-(|x_d-a|+|y_d|+a)E}-e^{-(|x_d|+|y_d-a|+a)E}
+e^{-(|x_d-a|+|y_d-a|)E}}{2E(1-e^{-2aE})}
\end{equation}
where $E=\sqrt{w_n^2+\mathbf{k}_\parallel^2+m^2}$.

We apply now the previous approach to the calculation of the first-order
correction to the free energy (in fact, to the force) and the self-energy
function:

\subsubsection{First-order correction to the free energy}
We want to evaluate thermal corrections to Casimir energy in the interacting 
theory, in particular, to understand how its divergences should be dealt
with. To automatically subtract $a$-independent contributions, we work with
the derivative of the free energy with respect to $a$. Obviously, this is a
force, and it has
the same amount of information as an energy which has been subtracted to
avoid $a$-independent infinities. Indeed, it can be integrated over any
finite range of distances to obtain de energy difference.

Using $F_{cas}$ to denote these derivatives, we have:
\begin{equation}
F_\textrm{cas}\,\,=\,\,F_\textrm{cas}^{(0)}+F_\textrm{cas}^{(1)}\,,
\end{equation}
Where the term $F_\textrm{cas}^{(1)}$ is the first-order correction to free
Casimir force $F_\textrm{cas}^{(0)}$.  There is a term where only
$\Delta(x,y)$ appears; this can be renormalized as usual in
finite-temperature field theory, by a zero-temperature subtraction plus the
inclusion of the first-order (temperature-dependent) mass counterterm
contribution~\cite{Kapusta:2006pm}. 

Then, keeping temperature {\em and} $a$ dependent terms only, we see that 
\begin{equation}
F_\textrm{cas}^{(1)}\,=\,F_{\textrm{cas},\,\Delta \,M}^{(1)}+F_{\textrm{cas},\,M\,M}^{(1)}\,\,,
\end{equation}
where:
\begin{equation}
F_{\textrm{cas},\,\Delta \,M}^{(1)}\,=\,-\frac{\lambda}{2 \beta}
\,\Delta^T(x,x)\int d^{d+1}x \,\frac{\partial M(x,x)}{\partial a}
\end{equation}
and
\begin{equation}
F_{\textrm{cas},\,M\,M}^{(1)}\,\,=\frac{\lambda}{4 \beta} \int d^{d+1}x \,M(x,x) \frac{\partial M(x,x)}{\partial a}\,\,.
\end{equation}
Using the notation $f_{\Delta,M}$ and $f_{M\,M}$ for the corresponding area
densities (we omit super and subscripts):  
\begin{equation}
f_{\Delta,M}\,=\,-\frac{\lambda}{2 \beta} \Delta^T(0) \sum_n \int
\frac{d^{d-1}\mathbf{k_\parallel}}{(2 \pi)^{d-1}}
\int_{-\infty}^{\infty}dx_d \frac{\partial
\widetilde{M}(\mathbf{k_\parallel},w_n,x_d)}{\partial a}\,\,,
\end{equation}
and
\begin{eqnarray}
f_{M,M}&=&\frac{\lambda}{4 \beta^2} \sum_{n,l} \int \frac{d^{d-1} 
\mathbf{k_\parallel}}{(2 \pi)^{d-1}}
\int \frac{d^{d-1}\mathbf{p_\parallel}}{(2 \pi)^{d-1}}  
\int_{-\infty}^{\infty}dx_d \nonumber\\
&\times& \widetilde{M}(\mathbf{k_\parallel},w_n,x_d) 
\frac{\partial \widetilde{M}(\mathbf{p_\parallel},w_l,x_d)}{\partial a}\;,
\end{eqnarray}
and the first-order correction to the Casimir force will be given by:
\begin{equation}
\frac{F_\textrm{cas}^{(1)}}{A_{d-1}}\,=\,f_{\Delta,M}+f_{M,M}\,\,.
\end{equation}

The term $\widetilde{M}(w_n,\mathbf{k}_\parallel;x_d)$ can be obtained from
(\ref{eq:Mpropagfourier}); after changing variables from $x\,\,
\to \,\,x+\frac{a}{2}$ (to obtain a symmetrized form), we perform an 
integration over $d+1$-dimensional spacetime, obtaining:
\begin{equation} 
f_{\Delta,M}\,=\,-\frac{\lambda}{4 \beta} \Delta^T(0) \sum_n \int
\frac{d^{d-1}\mathbf{k_\parallel}}{(2 \pi)^{d-1}}
\left(\frac{1}{2E_k}+\frac{a}{2}\,\textrm{csch}^2(aE_k)-\frac{\textrm{coth}(aE_k)}{2E_k}
\right)\;. 
\end{equation}
The $T$-dependent function $\Delta^T(0)$ is given by
\begin{equation}
\Delta^T(0)=\int \frac{d^d \mathbf{p}}{(2\pi)^d} \frac{1}{w} n_B(\beta,w)\,\,,
\end{equation}
where $n_B(\beta,\omega)$ is the Bose distribution function, with 
$\omega=\sqrt{\mathbf{p}^2+m^2}$.

$f_{\Delta,M}$ may be expressed in yet another way, using the
definition of the  Bose distribution function but with $\beta$ replaced by $2a$ 
while $\omega$ is replaced by $E_k=\sqrt{\omega_n^2+\mathbf{k}_\parallel^2+m^2}$, 
i.e., $n_B(\beta,\omega)$
is replaced by 
\begin{equation} 
n_k(2a,E_k) \equiv \frac{1}{e^{2a\, E_k}-1} \;.
\end{equation}
Then 
\begin{equation}
f_{\Delta,M}\,=\,\frac{\lambda}{4 \beta} \Delta^T(0) \sum_n 
\int \frac{d^{d-1}\mathbf{k_\parallel}}{(2 \pi)^{d-1}}
\left[\frac{n_k}{E_k}-2a \,n_k\,(1+n_k) \right]\,\,,
\end{equation}
which is clearly a convergent quantity.

For the remaining term, we proceed in a similar way, where now the 
$p$ represent $(\omega_l,\mathbf{p_\parallel})$ 
\begin{equation}
f_{M,M}\,=\,\frac{\lambda}{4 \beta^2} \sum_{n,l} \int \frac{d^{d-1}\mathbf{k_\parallel}}{(2 \pi)^{d-1}}
\int \frac{d^{d-1}\mathbf{p_\parallel}}{(2 \pi)^{d-1}} f(k,p)\,,
\end{equation}
where
\begin{eqnarray}
f(k,p) &=&
\left(n_p+1\right)\left(\frac{n_p}{E_k^2}-\frac{n_k
n_p}{E_k/2a}\right)\nonumber\\
&+&\left(n_p+\frac{1}{2}\right)\left(\frac{n_k}{E_k
E_p}-\frac{E_k\,n_p-E_p\,n_k}{E_k(E_k^2-E_p^2)}\right)-\frac{n_p}{2E_k(E_k+E_p)}
\, 
\end{eqnarray}
which is also convergent.

\subsubsection{First-order correction to the self-energy}
We conclude the application of this method with the calculation of the tadpole 
diagram contribution to the self-energy. There is here a surface divergence 
due to the $M(x;x)$ contribution:
\begin{equation}\label{eq:selener}
\Pi=\frac{\lambda}{2}(\Delta(x,x)- M(x,x))\,.
\end{equation}
As before, it is convenient to separate from the self-energy the
contribution which is present even when the mirrors are infinitely distant;
namely,
\begin{equation}
\Pi=\Pi_{free}+\Pi_{mir}\,,
\end{equation}
where $\Pi_{free}=\frac{\lambda}{2} \Delta(x,x)$ and $\Pi_{mir}=-
\frac{\lambda}{2} M(x,x)$ which contain the contribution to the selfenergy
coming from the Dirichlet boundary conditions. The term
$\Pi_{free}$ has UV and IR divergences, which are usually analyzed in
standard finite temperature calculations.

On the other hand, the $\Pi_{mir}$ term has no UV divergences, although it
is IR divergent in low dimensions.  Moreover, it presents surface
divergences when the $x_d$ coordinate tends to $x_d=0$ or $x_d=a$. We will
classify these divergences according to the dimension of the theory.

When $d=1$ and $m=0$, we obtain,
\begin{equation}
\Pi_{mir}=-\frac{\lambda}{2 \beta} \sum_n \frac{e^{-2|x|E}-2e^{-(|x|+|x-a|+a)E}+e^{-2|x-a|E} }{2E(1-e^{-2aE})} ,
\end{equation}
where $E=|\omega_n|$. Then we see that
\begin{displaymath}
\Pi_{mir} = -\frac{\lambda}{4 \pi} \sum_{n=1}^{\infty} \frac{1}{n} \times \left\{ \begin{array}{ll}
e^{\gamma s n} & \textrm{if $s<0$}\\
\\
\frac{cosh((s-1/2)\gamma n)-e^{-\gamma n /2}}{sinh (\gamma n/2)} & \textrm{if $0<s<1$}\\
\\
e^{-\gamma(s-1)n} & \textrm{if $s>1$}
  \end{array} \right.
\end{displaymath}

where $\gamma=4\pi a/\beta$ and $s=x/a$. The sum for $s<0$ and $s>1$ can be
performed. For $s<0$, the selfenergy is proportional to $\ln (1-exp(\gamma
s))$, which diverges for $s \to 0$, regardless of the temperature.

For $d\geq 2$, it is convenient to separate the self-energy part which
into two pieces, one coming from $n=0$, and the other coming
from the remaining modes:
\begin{equation}
\Pi_{mir}=\Pi_{mir}^{n=0}+\Pi_{mir}^{n \neq 0}\,,
\end{equation}
and 
\begin{displaymath}
\Pi_{mirr}^{n=0} = -\frac{\lambda}{4 \pi \beta} \int_{0}^{\infty} \, dp \, \frac{1}{p} \times \left\{ \begin{array}{ll}
e^{2s\,p} & \textrm{if $s<0$}\\
\\
\frac{cosh((2s-1)p)-e^{-p}}{sinh (p)} & \textrm{if $0<s<1$}\\
\\
e^{-2(s-1)p} & \textrm{if $s>1$}
  \end{array} \right.
\end{displaymath}

\begin{displaymath}
\Pi_{mirr}^{n \neq 0} = -\frac{\lambda}{2 \pi \beta} \sum_{n=1}^{\infty}  \times \left\{ \begin{array}{ll}
K_0 (-n \gamma s) & \textrm{if $s<0$}\\
\\
\int_1^{\infty} \frac{1}{\sqrt{p^2-1}}\frac{cosh((s-1/2)\gamma n p)-e^{-\gamma n p/2}}{sinh (\gamma n p/2)} & \textrm{if $0<s<1$}\\
\\
K_0 (n \gamma (s-1)) & \textrm{if $s>1$}
  \end{array} \right. \;.
\end{displaymath}

Again, the zero mode contribution $\Pi_{mirr}^{n=0}$ has an IR divergent
behaviour for low momenta, while there is a divergence on the mirrors.
For instance, for $s<0$ the selfenergy is proportional to  $log(-2s)$.

The $\Pi_{mirr}^{n \neq 0}$ term can be analized in the high and low
temperature limits, i.e. $\gamma \gg 1$ and $\gamma \ll 1$, respectively. 
The divergence of the function $K_0 (x)$ is
logarithmic in $x=0$, thus surface divergences also are logarithmic.
Moreover, we see that $\Pi_{mir}^{n = 0} \gg \Pi_{mir}^{n \neq 0}$.

For $d>2$,  the IR divergences are absent,  and the surface diverces appear
explicity.  The two contributions  are given by:
\begin{displaymath}
\Pi_{mirr}^{n=0} = -\frac{\lambda \,\,\Gamma(\frac{d}{2}-1)}{\beta \,2^{d+1} \pi^{\frac{d}{2}} a^{d-2}}  \times \left\{ \begin{array}{ll}
\frac{1}{s^{d-2}} & \textrm{if $s<0$}\\
\\
\zeta(d-2,1-s)+\zeta(d-2,s)-2\zeta(d-2) & \textrm{if $0<s<1$}\\
\\
\frac{1}{(s-1)^{d-2}} & \textrm{if $s>1$}
  \end{array} \right.
\end{displaymath}

\begin{displaymath}
\Pi_{mirr}^{n \neq 0} = -\frac{\lambda \, \pi^{\frac{d-3}{2}}}{2 \beta^{d-1} \Gamma(\frac{d-1}{2})} \sum_{n=1}^\infty \int_1^{\infty}dp\, n^{d-2} (p^2-1)^{\frac{d-3}{2}}   \left\{ \begin{array}{ll}
e^{\gamma n s\, p} & \textrm{if $s<0$}\\
\\
\frac{cosh((s-\frac{1}{2})\gamma n p)-e^{-\gamma n \frac{p}{2}}}{sinh (\gamma n \frac{p}{2})} & \textrm{if $0<s<1$}\\
\\
e^{-\gamma n (s-1)p}& \textrm{if $s>1$}
  \end{array} \right.
\end{displaymath}

The integral for $s<0$ or $s>1$ can be performed
\begin{displaymath}
\Pi_{mirr}^{n \neq 0} = -\frac{\lambda}{\beta^{\frac{d}{2}}}\frac{1}{(2a)^{\frac{d}{2}-1}} \sum_{n=1}^\infty
n^{\frac{d}{2}-1} \times \left\{ \begin{array}{ll}
(-1)^{\frac{d}{2}-1}\frac{K_{\frac{d}{2}-1}(-\gamma n s)}{s^{\frac{d}{2}-1}} & \textrm{if $s<0$}\\
\\
\frac{K_{\frac{d}{2}-1}(\gamma n (s-1))}{s^{\frac{d}{2}-1}}& \textrm{if $s>1$}
  \end{array} \right.
\end{displaymath}
In the case of $d=3$ dimension we have a simple expression
\begin{displaymath}
\Pi_{mirr}^{n \neq 0} = -\frac{\lambda}{8 \pi a \beta } \times \left\{ \begin{array}{ll}
-\frac{1}{s (e^{-\gamma s}-1)} & \textrm{if $s<0$}\\
\\
\frac{1}{(s-1) (e^{\gamma (s-1)}-1)}& \textrm{if $s>1$}
  \end{array} \right.
\end{displaymath}

By the above results we conclude that the surface divergences of
the $\Pi_{mirr}^{n = 0}$ term are polynomial: $\sim s^{-(d-2)}$,
whereas the term $\Pi_{mirr}^{n \neq 0}$ which has the sum of non-zero modes
presents a more severe divergence proportional to
$s^{-(d-1)}$. This has just been shown explicity for the $d=3$ case.

In figures 1, 2, 3 and 4, we plot the self-energy for 1, 2, and 3
dimensions, for different values of the parameters.
\subsection{Non-perturbative approach}\label{ssec:nonpert}
Let us conclude this section by considering a second, alternative procedure
to the one just explained. It amounts to integrating out, albeit formally,
the scalar field {\em before\/} the perturbative expansion. Indeed,
introducing, from the very beginning, the auxiliary fields used to impose
the Dirichlet conditions, we see that the partition function becomes
\begin{equation}
{\mathcal Z}(\beta,a) \;=\;\int {\mathcal D}\xi_1 {\mathcal D}\xi_2 \; 
e^{- {\mathcal W}[i J_p]} \; ,
\end{equation}
where ${\mathcal W}$ denotes the generating functional of connected
correlation functions, at finite temperature, and with an unconstrained (no
Dirichlet conditions) scalar field integration measure:
\begin{equation}
e^{- {\mathcal W}[J]} \;\equiv\; \int {\mathcal D}\varphi \, e^{-S[\varphi]
+ \int d^{d+1}x J(x) \varphi(x) } \;.
\end{equation}
Of course, the arbitrary current $J$ in the definition above must be
replaced by $J_p$, which does depend on the auxiliary fields and on $a$,
the distance between the two mirrors. Besides, we assume that, in the
course of evaluating ${\mathcal W}$, a renormalization procedure has been
used to make sense of the possible infinites, in the usual way.

To proceed, we recall  that ${\mathcal W}[J]$ does have a functional
expansion:
\begin{equation}
{\mathcal W}[J] \;=\; {\mathcal W}_0 \, + \, \frac{1}{2} \int d^{d+1}x \int
d^{d+1}y \; {\mathcal W}_2(x,y) J(x) J(y) \,+\, \ldots  
\end{equation}
where ${\mathcal W}_k$ correponds to the connected $k$-point correlation
functions.  Odd terms are, for the quartic perturbation, absent from the expansion.
Then one may invoque some approximation that allows one to truncate the
functional expansion. Of course, it cannot be a naive perturbative
expansion in the coupling constant; one should rather use, for example, a
mean field or large-$N$ expansion. Then the leading term will be just the
quadratic one:
\begin{equation}
{\mathcal Z}(\beta,a) \;\sim\; {\mathcal Z}(\beta) \,\times \,
{\mathcal Z}_q(\beta,a)
\end{equation}
where ${\mathcal Z}(\beta)= e^{{\mathcal W}_0}$ is the thermal partition function in the absence
of mirrors, while
\begin{equation}
{\mathcal Z}_q(\beta,a) \;=\; \int {\mathcal D}\xi_1 {\mathcal D}\xi_2 \; 
e^{- \frac{1}{2} \int d^{d+1}x \int
d^{d+1}y \; {\mathcal W}_2(x,y) J_p(x) J_p(y)} \;,
\end{equation}
depends on ${\mathcal W}_2(x,y)$, the full renormalized thermal propagator.  
Using the explicit form of $J_p$, we see that the integral is a Gaussian:
\begin{equation}\label{eq:zq1}
{\mathcal Z}_q(\beta,a) \;=\; \int {\mathcal D}\xi_1 {\mathcal D}\xi_2 \; e^{-S_q(\xi)}
\end{equation}
where
\begin{equation}
	S_q(\xi) \;=\; \frac{1}{2} \,\int d^dx_\parallel  d^dy_\parallel  
	\xi_a(x_\parallel) \Omega^{q}_{ab}(x_\parallel,y_\parallel)
	\xi_b(y_\parallel) \;,
\end{equation}
and 
\begin{equation}
\Omega^q(x_{\parallel};y_{\parallel})=\left[ \begin{array}{cc}
D(x_{\parallel},0;y_{\parallel},0) &
D(x_{\parallel},0;y_{\parallel},a) \\
D(x_{\parallel},a;y_{\parallel},0) &
D(x_{\parallel},a;y_{\parallel},a) \\
\end{array} \right] \;,
\end{equation}
where $D$ is the {\em full\/} imaginary-time propagator. 
It may be written quite generally as follows:
\begin{equation}
D(\tau_x,\mathbf{x};\tau_y,\mathbf{y})=\frac{1}{\beta}\sum_n \int
\frac{d^d\mathbf{k}}{(2\pi)^d} e^{i\omega_n(\tau_x\ - \tau_y)
+i\mathbf{k}(\mathbf{x}-\mathbf{y})}\,
\widetilde{D}(\omega_n,\mathbf{k})\;,
\end{equation}
with \mbox{$\tilde{D}(\omega_n,{\mathbf k}_\parallel)$} a
generally complicated function of its arguments. However, we note that some
non perturbative corrections do produce a simple result. For example,
considering the IR resummed version of the massless scalar field, yields 
an expression of the form:
\begin{equation}
\tilde{D}(\omega_n,\mathbf{k}) \,=\, \big[ \omega_n^2 + {\mathbf
k}_\parallel^2 + \Pi_\beta(T) \big]^{-1}
\end{equation}
where $\Pi_\beta(T)$ is the thermal mass. For example, the first two non-trivial
contributions correspond to a term which is linear in $\lambda$ plus a
non-analytic term:
\begin{equation}
\Pi_\beta(T) \,=\, \frac{\lambda T^2}{24} \big[ 
1 \,-\, 3 ( \frac{\lambda}{24 \pi^2} )^{\frac{1}{2}} \,+\, \ldots \big] 
\;. 
\end{equation} 
Upon insertion of this expression, we see that the corresponding
contribution to the Casimir free energy becomes:
\begin{equation}
{\mathcal F}_c(\beta,a) \;=\; \frac{1}{2 \beta} \int
\frac{d^{d-1}{\mathbf k}_\parallel}{(2\pi)^{d-1}}  \sum_{n
=-\infty}^{+\infty} \ln \Big[ 1 \,-\, e^{- 2 a  \sqrt{\omega_n^2 + {\mathbf
k}_\parallel^2 + \Pi_\beta(T)}} \Big] \;,
\end{equation}
where the $a \to \infty$ contribution has already been subtracted.

\section{Conclusions}\label{sec:concl}
We have obtained exact results for the free energy in a Casimir system 
in $1+1$ dimensions and well as low and high temperature expansions for 
$d>1$, based on a duality relation between inverse temperatures and
distances between mirrors, in $d+1$ spacetime dimensions.

For the interacting theory, we have derived two different approaches.  In
the first approach, perturbative expansion has been used to obtain the first
order corrections to thermal Casimir force, showing that it is finite once
the standar renormalization of thermal field theory in the absence of
mirrors is performed. 

On the other hand, we also studied the selfenergies, showing that they have 
surface divergences in their coordinate dependence, and that those
divergences are of a polynomial type, with a degree which depends on the
dimension.

Finally, we have shown that a variation in the order used to integrate the
auxiliary fields yields a different method, whereby the one particle
irreducible functions naturally appear.

\section*{Acknowledgements}
C.C.T and C.D.F. thank CONICET, ANPCyT and UNCuyo for financial support.

\newpage
\begin{picture}(0,0)%
\includegraphics[width=0.8\textwidth]{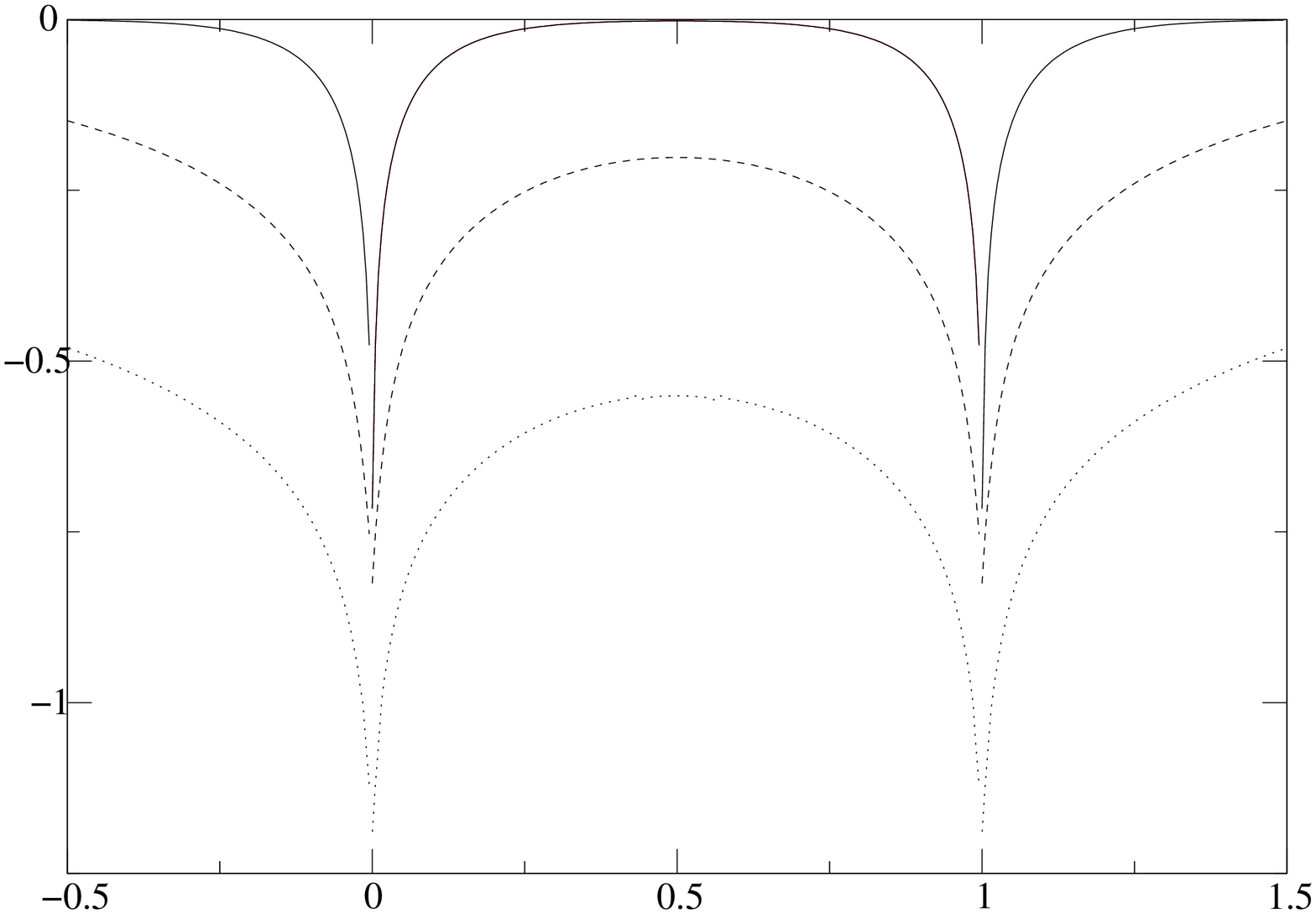}
\end{picture}%
\setlength{\unitlength}{3947sp}%
\begingroup\makeatletter\ifx\SetFigFont\undefined%
\gdef\SetFigFont#1#2#3#4#5{%
  \reset@font\fontsize{#1}{#2pt}%
  \fontfamily{#3}\fontseries{#4}\fontshape{#5}%
  \selectfont}%
\fi\endgroup%
\begin{picture}(4064,2611)(-14,-12500)
\put(-20,-10500){\makebox(0,0)[lb]{\smash{{\SetFigFont{14}{16.4}{\rmdefault}{\mddefault}{\updefault}{\color[rgb]{0,0,0}$\frac{\lambda}{4 \pi}$}%
}}}}
\put(2550,-12256){\makebox(0,0)[lb]{\smash{{\SetFigFont{12}{14.4}{\rmdefault}{\mddefault}{\updefault}{\color[rgb]{0,0,0}$s$}%
}}}}
\end{picture}%

\newpage

\begin{picture}(0,0)%
\includegraphics{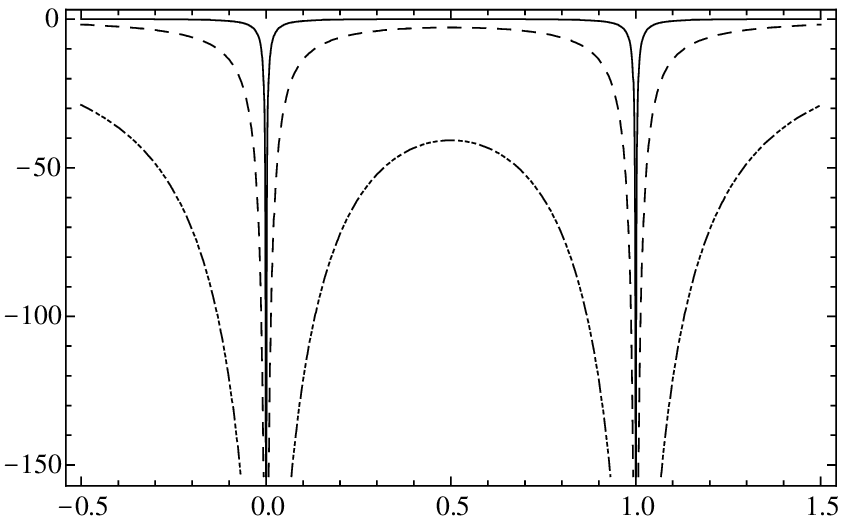}%
\end{picture}%
\setlength{\unitlength}{3947sp}%
\begingroup\makeatletter\ifx\SetFigFont\undefined%
\gdef\SetFigFont#1#2#3#4#5{%
  \reset@font\fontsize{#1}{#2pt}%
  \fontfamily{#3}\fontseries{#4}\fontshape{#5}%
  \selectfont}%
\fi\endgroup%
\begin{picture}(4064,2611)(-14,-12500)
\put(-100,-11086){\makebox(0,0)[lb]{\smash{{\SetFigFont{14}{16.4}{\rmdefault}{\mddefault}{\updefault}{\color[rgb]{0,0,0}$\frac{\lambda}{2 \pi \beta}$}%
}}}}
\put(2100,-12536){\makebox(0,0)[lb]{\smash{{\SetFigFont{12}{14.4}{\rmdefault}{\mddefault}{\updefault}{\color[rgb]{0,0,0}$s$}%
}}}}
\end{picture}%

\newpage

\begin{picture}(0,0)%
\includegraphics{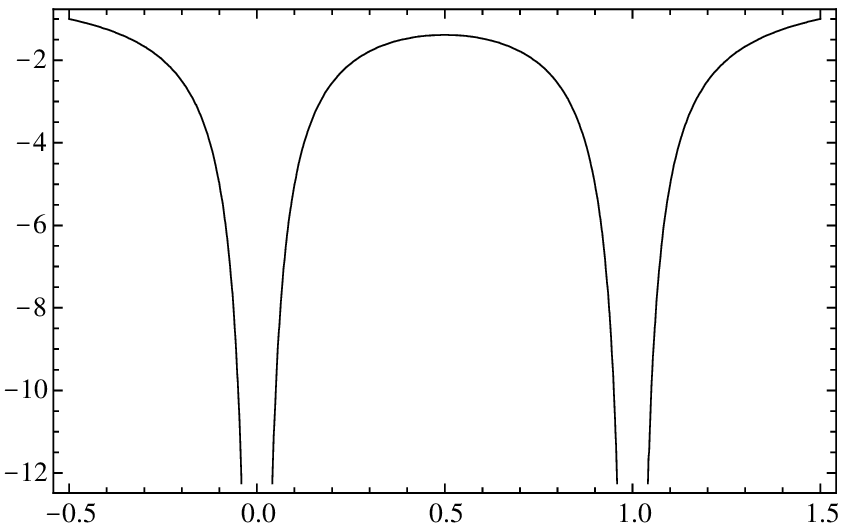}%
\end{picture}%
\setlength{\unitlength}{3947sp}%
\begingroup\makeatletter\ifx\SetFigFont\undefined%
\gdef\SetFigFont#1#2#3#4#5{%
  \reset@font\fontsize{#1}{#2pt}%
  \fontfamily{#3}\fontseries{#4}\fontshape{#5}%
  \selectfont}%
\fi\endgroup%
\begin{picture}(4063,2714)(-14,-12575)
\put(  -300,-11161){\makebox(0,0)[lb]{\smash{{\SetFigFont{14}{16.4}{\rmdefault}{\mddefault}{\updefault}{\color[rgb]{0,0,0}$\frac{\lambda}{8 \pi a \beta}$}%
}}}}
\put(2130,-12511){\makebox(0,0)[lb]{\smash{{\SetFigFont{12}{14.4}{\rmdefault}{\mddefault}{\updefault}{\color[rgb]{0,0,0}$s$}%
}}}}
\end{picture}%

\newpage

\begin{picture}(0,0)%
\includegraphics{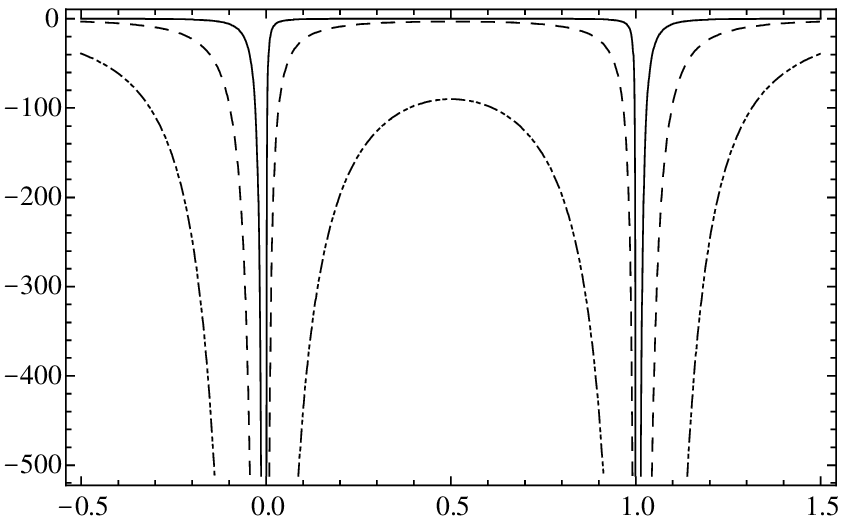}%
\end{picture}%
\setlength{\unitlength}{3947sp}%
\begingroup\makeatletter\ifx\SetFigFont\undefined%
\gdef\SetFigFont#1#2#3#4#5{%
  \reset@font\fontsize{#1}{#2pt}%
  \fontfamily{#3}\fontseries{#4}\fontshape{#5}%
  \selectfont}%
\fi\endgroup%
\begin{picture}(4064,2536)(-14,-12425)
\put(2060,-12500){\makebox(0,0)[lb]{\smash{{\SetFigFont{12}{14.4}{\rmdefault}{\mddefault}{\updefault}{\color[rgb]{0,0,0}$s$}%
}}}}
\put(  -300,-11086){\makebox(0,0)[lb]{\smash{{\SetFigFont{14}{16.4}{\rmdefault}{\mddefault}{\updefault}{\color[rgb]{0,0,0}$\frac{\lambda}{8 \pi a \beta}$}%
}}}}
\end{picture}%

\newpage
\begin{center}
	{\large \bf \sc Figure Captions}
\end{center}
\begin{itemize}
	\item {\bf Figure 1:} Self-energy in $1+1$ dimensions, for three
		values of $\gamma \equiv  4 \pi a / \beta$: upper curve:
		$\gamma=10$; middle curve: $\gamma=1$ and lower curve:
		$\gamma=0.1$.
	\item {\bf Figure 2:} Self-energy in $2+1$ dimensions, for
		three values of $\gamma \equiv  4 \pi a / \beta$: upper
		curve: $\gamma=10$; middle curve: $\gamma=1$ and lower
		curve: $\gamma=0.1$.
	\item {\bf Figure 3:} Self-energy in $3+1$ dimensions, zero mode
		contribution.
	\item {\bf Figure 4:}  Self-energy in $3+1$ dimensions, for
		three values of $\gamma \equiv  4 \pi a / \beta$: upper
		curve: $\gamma=10$; middle curve: $\gamma=1$ and lower
		curve: $\gamma=0.1$.

\end{itemize}

\end{document}